\documentclass[12pt]{iopart}
\usepackage{graphicx}
\begin{document}

\title[]{Detector configuration of KAGRA - the Japanese cryogenic gravitational-wave detector -}

\author{Kentaro Somiya (for the KAGRA Collaboration)}
\address{Graduate School of Science and Technology, Tokyo Institute of Technology,\\ 2-12-1 Oh-okayama, Meguro, Tokyo, 152-8551, Japan}
\ead{somiya@phys.titech.ac.jp}
\begin{abstract}
Construction of the Japanese second-generation gravitational-wave detector KAGRA (previously called LCGT) has been started. In the next $6\sim7$ years, we will be able to observe the space-time ripple from faraway galaxies. KAGRA is equipped with the latest advanced technologies. The entire 3-km long detector is located in the underground to be isolated from the seismic motion, the core optics are cooled down to 20~K to reduce thermal fluctuations, and quantum non-demolition techniques are used to decrease quantum noise. In this paper, we introduce the detector configuration of KAGRA; its design, strategy, and downselection of parameters.
\end{abstract}

\maketitle

\section{Introduction}

KAGRA (previously called LCGT for Large-scale Cryogenic Gravitational-wave Telescope) is a Japanese 3-km optical interferometer currently under construction in the Kamioka mine. The aim of the detector is frequent observation of gravitational waves from faraway galaxies and to obtain unique information on the universe. The observation will be performed with other detectors in the US (Advanced LIGO~\cite{LIGO}) and Europe (Advanced Virgo~\cite{Virgo} and GEO-HF~\cite{GEO}), which are now being upgraded after several-year observations in the initial configurations. KAGRA and the other three detectors are called second-generation gravitational-wave detectors.

There are two unique features in the detector configuration of KAGRA. One is that the entire detector is constructed in the underground. Both seismic noise and gravity gradient noise are low in the underground. In addition, the low rms motion relaxes the requirement on the interferometer control and reduces electro-magnetic noise. The other special feature is the cryogenic operation of the interferometer. The sapphire test masses are cooled down to 20~K and mirror thermal noise is lower than that of room-temperature detectors. Another benefit of the cryogenic operation is that the mirror causes almost no thermal lensing effect, which is one of the biggest issues for the room-temperature detectors, for the high thermal conductivity of sapphire at 20~K.

The unique features of KAGRA come along with unique issues. The baseline length is limited by the size of the mountain. The floor is tilted for the water drainage system. The mirror is made of sapphire instead of classic fused silica. The laser power is limited by the amount of the absorbed heat that can be extracted through the suspension fibers. The detector configuration of KAGRA is designed with these issues fully considered. Some of our knowledge of the underground cryogenic interferometer will be useful for a third-generation gravitational-wave detector, Einstein Telescope~\cite{ET} for example, which is planned to be built underground and operated in a cryogenic temperature.

\begin{figure}[htbp]
	\begin{center}
		\includegraphics[width=12cm]{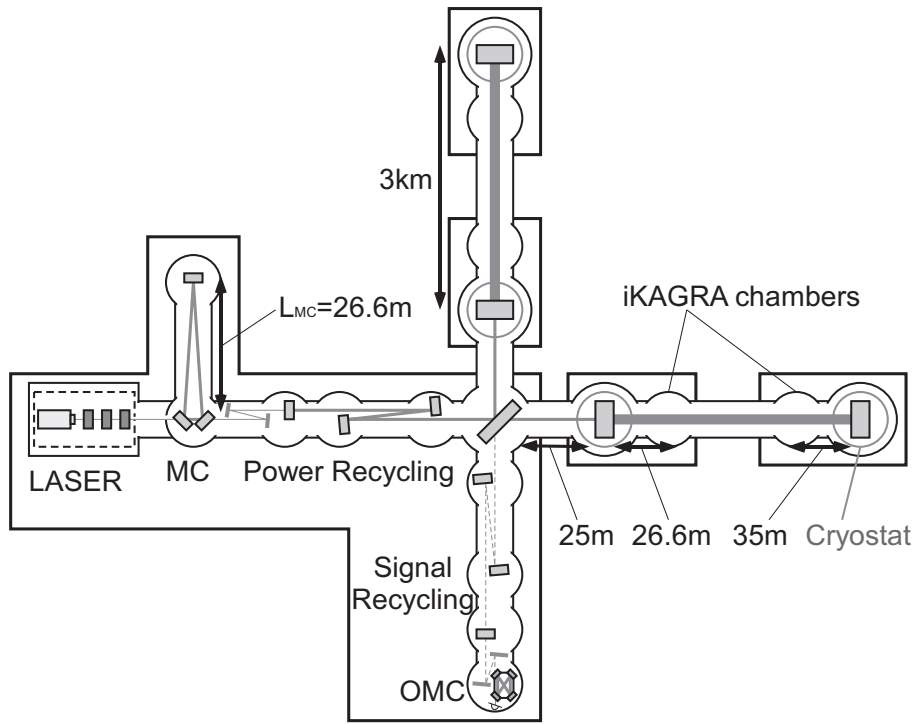}
	\caption{Schematic view of KAGRA. The test masses are installed in the cryostats and are isolated from room-temperature optics by more than 20~m. The mirrors for iKAGRA will be installed in vacuum chambers next to the cryostats for the smooth transition to bKAGRA. Here MC/OMC stand for mode-cleaner/output mode-cleaner.}
	\label{fig:IFO}
	\end{center}
\end{figure}

Figure~\ref{fig:IFO} shows the schematic view of the KAGRA detector. The optical configuration is a Michelson interferometer with the high-finesse arm cavities and the folded recycling cavities. The signal recycling cavity reduces the storage time of the signal fields resonating in the arm cavities in order to retain a broad bandwidth; this system is called resonant-sideband extraction (RSE)~\cite{Mizuno}. The four test-mass mirrors in the arm cavities are cooled down to 20~K to reduce thermal fluctuations. These mirrors are made of sapphire and suspended by sapphire fibers. The mass of the sapphire mirrors ranges between $23\sim30$~kg, depending on the availability. The laser power after the mode-cleaner ranges between $50\sim80$~W, depending on the absorption of the sapphire mirror; the less absorption, the more power we can inject. The power recycling gain is about 10 and the finesse of the arm cavity is 1550. The power reflectivity of the signal recycling mirror is 85~\% and the resonant condition of the signal recycling cavity is detuned by $\sim3.5$~deg to increase the detector sensitivity to the gravitational waves from the neutron-star binaries (the major parameters are listed in \ref{appA}). Our primary target source is the neutron-star binaries since the event rate can be well estimated and the waveform during the inspiral is predictable. In the end, the observable distance of the neutron-star binary inspirals with KAGRA ranges between $240\sim280$~Mpc, with which we will be able to detect $6\sim10$ neutron-star binary signals per year~{\cite{range1}\cite{range2}\cite{range3}}. Here the normal incidence of the wave is assumed in the calculation of inspiral range.

KAGRA project was approved by the Japanese government in 2010. The initial phase will be the room-temperature operation without the recycling cavities. This is called initial KAGRA or iKAGRA. A short-term observation of iKAGRA is planned in 2015. The test masses of iKAGRA are 10~kg silica mirrors provided by courtesy of LIGO. In the second phase, the test masses are replaced by the 23~kg (or 30~kg) sapphire mirrors to be cooled down. The suspension system is upgraded and the recycling mirrors are introduced. This is the baseline-design KAGRA or bKAGRA. Observations with bKAGRA will start in $2017\sim2018$.

\section{Estimated noise budget}

\begin{figure}[htbp]
	\begin{center}
		\includegraphics[width=12cm]{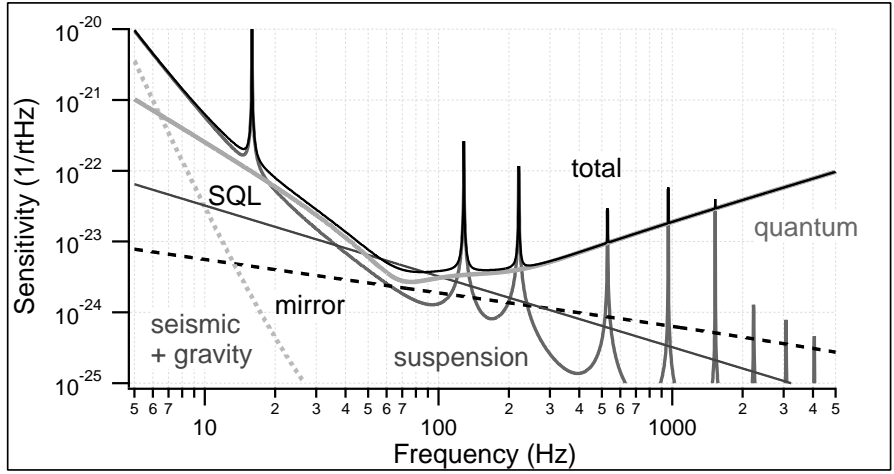}
	\caption{The estimated noise budget of KAGRA.}
	\label{fig:spectrum}
	\end{center}
\end{figure}

Figure~\ref{fig:spectrum} shows the estimated noise budget of KAGRA~\cite{official}. Hereafter we assume the safest parameter set, namely the 23~kg mass and the 50~W laser. The sensitivity is limited by quantum noise at most frequencies, and also suspension thermal noise is close to limit the sensitivity at low frequencies ($20\sim100$~Hz). This reflects quite well the characteristic of the detector. While mirror thermal noise is low for the cryogenic operation, shot noise is higher than other advanced detectors due to the low laser power for the cryogenic operation. The lighter mass increases quantum radiation pressure noise and suspension thermal noise. The fiber thickness is not determined by its strength but by the amount of the heat transferred from the test mass to the upper stages that are connected to the cryo-cooler. The thick fiber increases suspension thermal noise. The peak of suspension thermal noise at 130~Hz is the vertical-mode resonance and the peaks starting at 230~Hz are the violin modes. The peak at 16~Hz is the vertical resonance of the wire suspending the recoil mass.

We discuss more in detail about each noise curve in the following sections. Seismic noise and gravity gradient noise (GGN) are discussed in Sec.~\ref{sec:seismic}, thermal noise is discussed in Sec.~\ref{sec:thermal}, and quantum noise is discussed in Sec.~\ref{sec:QN}.

\section{Underground observatory}\label{sec:seismic}

\begin{figure}[htbp]
	\begin{center}
		\includegraphics[width=7cm]{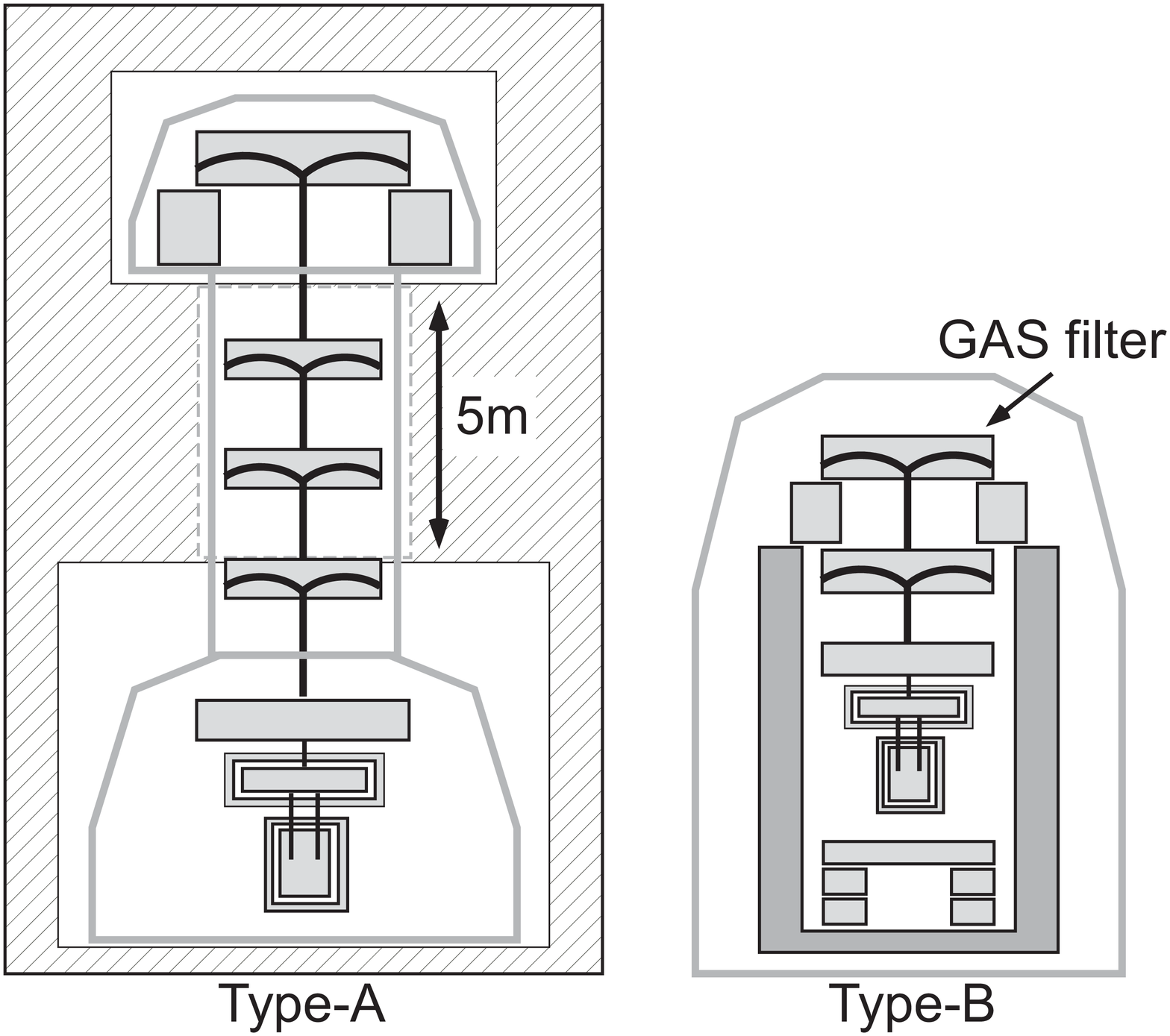}
\hspace{0.4cm}
		\includegraphics[width=7.5cm]{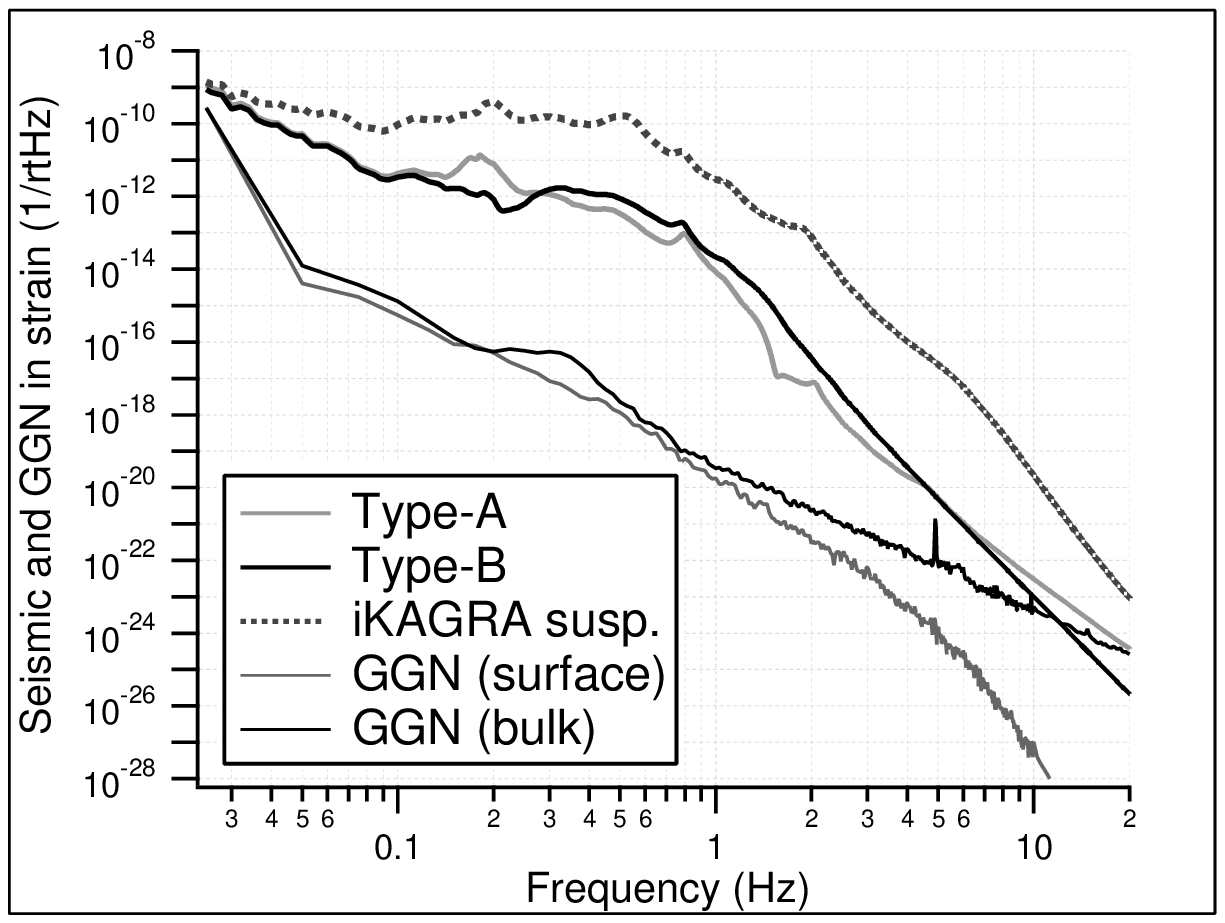}
	\caption{{\it Left}: Schematic view of the suspension systems. The type-A is for the test masses and the type-B is for the beamsplitter and the recycling mirrors. The table underneath the suspended mass in type-B is for auxiliary optics. {\it Right}: Seismic noise and gravity gradient noise (GGN) spectra.}
	\label{fig:VIS}
	\end{center}
\end{figure}

On the left side of Fig.~\ref{fig:VIS} are shown the schematic views of type-A and type-B suspensions. The type-A is a 2-story suspension with short and steady inverted pendula on the top floor, 4 geometric anti-spring (GAS) filters in the middle, and the triple pendulum at the bottom for the cryogenic test masses. The type-B is for the beamsplitter and the recycling mirrors. While the type-A suspension system has more GAS filters and the isolation ratio is better than type-B at low frequencies, vibrations of the cryo-cooler transferred through the heat link limit the noise performance above a few Hz. In the end the noise levels of type-A and type-B are nearly equal in the observation band.

The seismic motion of the KAGRA site in the Kamioka mine is $\sim100$ times lower than that of the TAMA site at $1\sim100$~Hz~\cite{LISM}. The temperature and the humidity fluctuations measured in the mine are 0.01~deg per day and 0.08~\% per day, respectively~\cite{LISM}.

On the right side of Fig.~\ref{fig:VIS} are shown the seismic noise spectra with the type-A and type-B suspensions together with the GGN spectra, which were simulated by courtesy of Jan Harms at Caltech using seismic data of the Kamioka region provided by a research group in NIKHEF~\cite{NIKHEF}. The test masses are located at least 200~m away from the surface of the mountain. It is an advantage of the underground observatory that GGN from the surface of the mountain is suppressed.

\section{Cryogenic operation}\label{sec:thermal}

\subsection{Heat transfer}

\begin{figure}[t]
	\begin{center}
		\includegraphics[width=11cm]{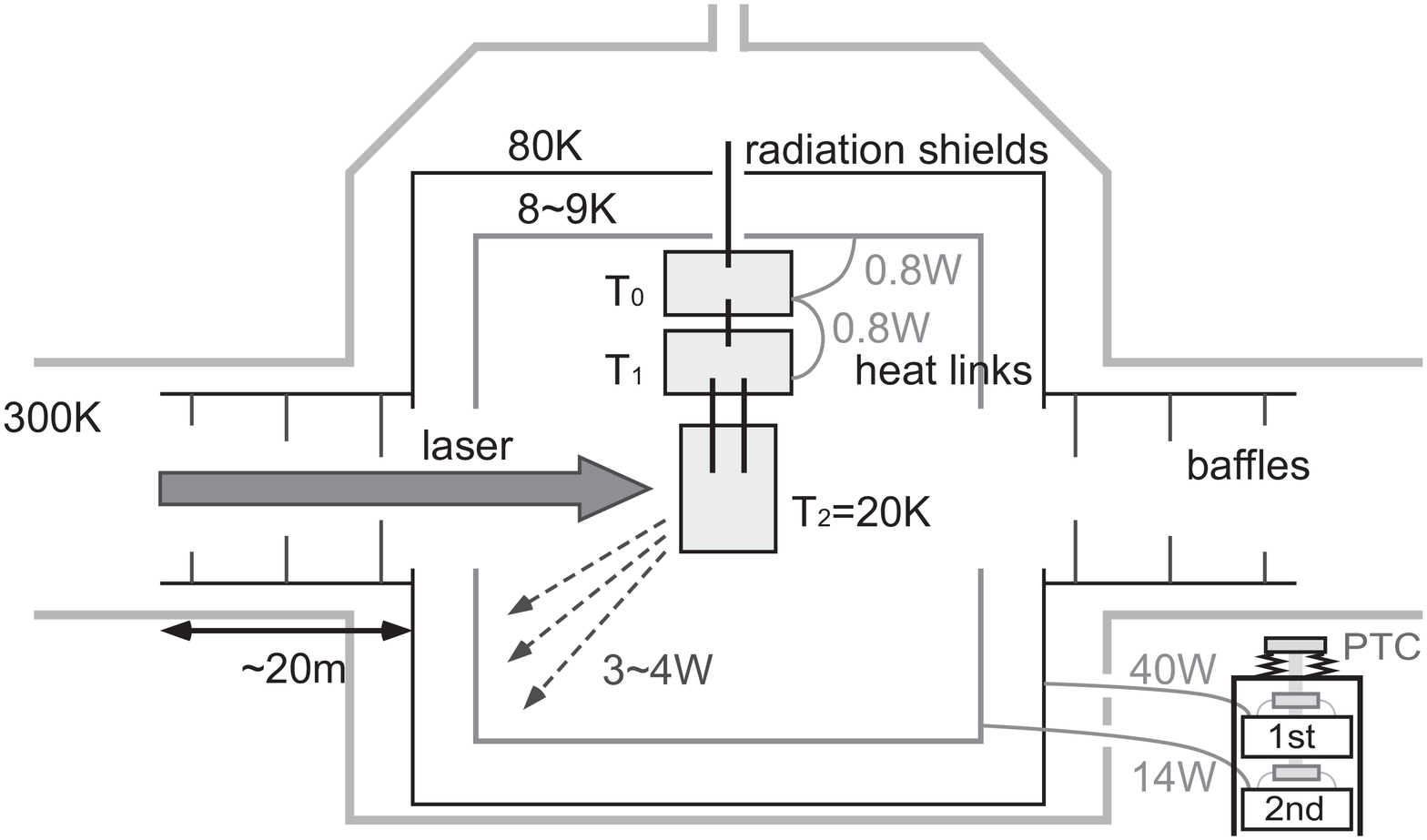}
	\caption{Schematic view of the cryostat. The heat absorbed from the laser light is transferred from the test mass to the upper stages and then extracted to the pulse-tube cryo-cooler (PTC). The radiation shield is extended along the vacuum duct over 20~m in order to reduce the heat from the 300~K radiation.}
	\label{fig:cryostat}
	\end{center}
\end{figure}

Figure~\ref{fig:cryostat} shows the schematic view of the cryogenic system that contains a sapphire test mass suspended by sapphire fibers in the cryostat vacuum chamber. The outer/inner radiation shield is connected to the 1st-stage/2nd-stage of the cryo-cooler through the pure-metal heat path and cooled down to 80~K/8~K. The two stages are connected to the pulse-tube cryo-cooler (PTC) through the flexible heat links and the PTC is on the vibration reduction stage~\cite{Tomaru}.

The last 3 stages of the suspension system are inside the radiation shields. The platform is connected to the inner shield through the 7 pure aluminum heat links of 1~mm diameter, and the intermediate mass suspended under the platform is connected to the platform through the 5 heat links of 3~mm diameter. The heat absorbed in the test mass is transferred to the intermediate mass through the suspension fiber and then transferred to the upper stages through the heat links.

The transferable heat from an object at temperature $T_2$ to an object at temperature $T_1$ is given by the following equation:
\begin{eqnarray}
K=\int_{T_1}^{T_2}\frac{\pi d_w^2}{4\ell_{sus}}N_w\kappa(d_w,\ T) dT\ ,
\end{eqnarray}
where $d_w$, $\ell_{sus}$, $N_w$, and $\kappa$ are diameter, length, number, and thermal conductivity of the fiber or the heat link, respectively. The thermal conductivity depends on the fiber diameter due to the surface scattering and also depends highly on the temperature. While the thermal conductivity can be calculated using the Debye model~\cite{Morelli}, we use the following simple equation that is derived from measurement results and is good below $\sim40$~K: $\kappa\simeq5270\times d_w(T/1\mathrm{K})^{2.24}$~(W/m/K).

With the full laser power injected, about 800~W light transmits through the input test mass (ITM) and 400~kW light is stored in the cavity. The absorption of the sapphire substrate ranges between $32\sim67$~ppm/cm in a recent measurement~\cite{Hirose}. The absorption of the silica/tantala doublet coatings measured at 20~K is $0.4\sim0.5$~ppm~\cite{Yama}. To be on a safer side, we assume 50~ppm/cm for the substrate and 1.0~ppm for the coatings. There is additional 200~mW heat introduced through the radiation from the apertures and the view ports. In total, about 1.2~W heat is absorbed by the test mass with the full power. There are point-like defects on the mirror surface and about 10~ppm of the light in the cavity is scattered around to contribute to increase the temperature of the inner radiation shield. The inner shield is coated with the diamond-like carbon (DLC) to reduce time to cool down the test masses before the operation. 
The absorptivity of the 1.0~$\mu\mathrm{m}$ DLC coating on the CP aluminum at 10~$\mu\mathrm{m}$ is as high as 0.41, so that the mirror can be cooled faster with the radiation from the mirror to the inner shield~\cite{Sakakibara}. On the other hand, the scattered light is also absorbed by the inner shield more effectively. Most of the scattered light, about $3\sim4$~W, is absorbed and the heat can increase the temperature of the inner shield to $\sim9$~K. All of these facts taken into account, the test mass temperature would become 23~K with the full laser power.

As is shown in the left panel of Fig.~\ref{fig:TN}, mirror thermal noise, especially substrate thermoelastic noise, increases with the mirror temperature. The inspiral range of the neutron-star binary with the full-power operation decreases from 250~Mpc to 230~Mpc when the mirror temperature increases from 20~K to 23~K. It is then rather better to decrease the laser power. The mirror temperature can be 20~K with about 60~\% of the laser power. Shot noise increases and thermal noise decreases. The inspiral range is then 240~Mpc.

An alternative solution for the increasing heat absorption would be to use thicker fibers. Replacing the 1.6~mm fiber by a 2.0~mm fiber, we could retain the mirror temperature of 20~K with the full laser power injected in return to an increase of suspension thermal noise. The inspiral range would be then slightly higher, but this strategy takes away the possibility of the sensitivity improvement in the case the absorption be lower than the estimate.

\subsection{Mirror thermal noise}

The mirror thermal noise curve in Fig.~\ref{fig:spectrum} is the sum of substrate Brownian noise, coating Brownian noise, and substrate thermoelastic noise. Coating thermo-optic noise is supposedly low at 20~K and is ignored here. 

The substrate Brownian noise and coating Brownian noise levels for each mirror are given by~{\cite{Levin}\cite{Harry}}
\begin{eqnarray}
&&S_\mathrm{sub}(\Omega)=\frac{4k_\mathrm{B}T}{\Omega}\frac{\phi_\mathrm{s}}{\sqrt{\pi}w_0}\times\frac{1-\nu_\mathrm{s}^2}{Y_s}\ ,\label{eq:TN1}\\
&&S_\mathrm{coa}(\Omega)=\frac{4k_\mathrm{B}T}{\Omega}\frac{d_c\phi_\mathrm{c}}{\pi w_0^2}\times\frac{Y_\mathrm{c}^2(1+\nu_\mathrm{s})^2(1-2\nu_\mathrm{s})^2+Y_\mathrm{s}^2(1+\nu_\mathrm{c})^2(1-2\nu_\mathrm{c})}{Y_\mathrm{s}^2Y_\mathrm{c}(1-\nu_\mathrm{c}^2)}\ ,\nonumber\\ \label{eq:TN2}
\end{eqnarray}
respectively. Here $k_\mathrm{B}$, $T$, $\Omega/2\pi$, $d_c$, $w_0$, $\nu_j$, $Y_j$, and $\phi_j$ are the Boltzmann constant, mirror temperature, frequency, coating thickness, beam radius, Poisson ratio, Young's modulus, and mechanical loss angle, respectively, with the subscript $j$ indicating substrate for $s$ and coatings for $c$. As for the coatings, the total thermal noise level is a sum of the noise level for silica layers and the noise level for tantala layers.

\begin{table}[htbp]
\begin{center}
\begin{tabular}{|c|c|p{2.3cm}|p{2.3cm}|p{2.3cm}|}
\hline
material&Young's mod.&substrate&silica layer&tantala layer\\ \hline
sapphire&400~GPa&2.3e-12&1.3e-11&6.4e-12\\ \hline
silica&72~GPa&1.3e-11&2.1e-11&2.3e-11\\ \hline
silicon&188~GPa&5.0e-12&1.4e-11&8.0e-12\\ \hline
\end{tabular}
\caption{The last fractions in Eqs.~(\ref{eq:TN1}) and (\ref{eq:TN2}) that consist of Young's modulus and Poisson ratio (in the unit of 1/Pa). The sapphire substrate has an advantage that these coefficients are low.}
\label{tab:coef}
\end{center}
\end{table}

KAGRA has two strong points and one weak point on the Brownian noise levels. One strong point is certainly the low temperature. The Brownian noise level is proportional to $T$. The other strong point is the high Young's modulus of sapphire. Table~\ref{tab:coef} shows the comparison of the last fractions in Eqs.~(\ref{eq:TN1}) and (\ref{eq:TN2}) for different substrate materials. The weak point is the mechanical loss of the coating materials at 20~K. Table~\ref{tab:loss} shows the mechanical losses of the coating layers measured in the University of Tokyo and the University of Glasgow. It is reported in Glasgow that the mechanical loss can be decreased by doping titanium or by choosing a proper annealing temperature. It has not yet been reported if both effects can be accumulated. The experiment in Tokyo does not see any effect of changing the annealing temperature but the mechanical loss is lower than what is measured in Glasgow. The experiment in Tokyo is planned to be restarted and more investigations will be done to achieve the goal.

\begin{table}[htbp]
\begin{center}
\begin{tabular}{|l|p{1.4cm}|p{1.4cm}|p{1.4cm}|}
\hline
&tantala&silica&average\\ \hline
290~K~\cite{LMA}&2.0e-4&5.0e-5&1.3e-4\\ \hline
20~K, measured in Tokyo~\cite{Yama}&--&--&5.0e-4\\ \hline
20~K, 600~$^\circ${\rm C} anneal~\cite{Glasgow1}&1.0e-3&5.0e-4&7.8e-4\\ \hline
20~K, 600~$^\circ${\rm C} anneal, Ti-doped~\cite{Glasgow1}&8.0e-4&5.0e-4&6.7e-4\\ \hline
20~K, 300~$^\circ${\rm C} anneal~\cite{Glasgow2}&6.0e-4&--&--\\ \hline
20~K, 300~$^\circ${\rm C} anneal, Ti-doped&--&--&--\\ \hline
goal for KAGRA&5.0e-4&3.0e-4&4.1e-4\\ \hline
\end{tabular}
\caption{Measured mechanical losses of the coating layers. The average loss is calculated by $(\sum Y_nd_n\phi_n)/(\sum Y_nd_n)$ with the Poisson ratios ignored~\cite{multi}.}
\label{tab:loss}
\end{center}
\end{table}

\begin{figure}[t]
	\begin{center}
		\includegraphics[width=7cm]{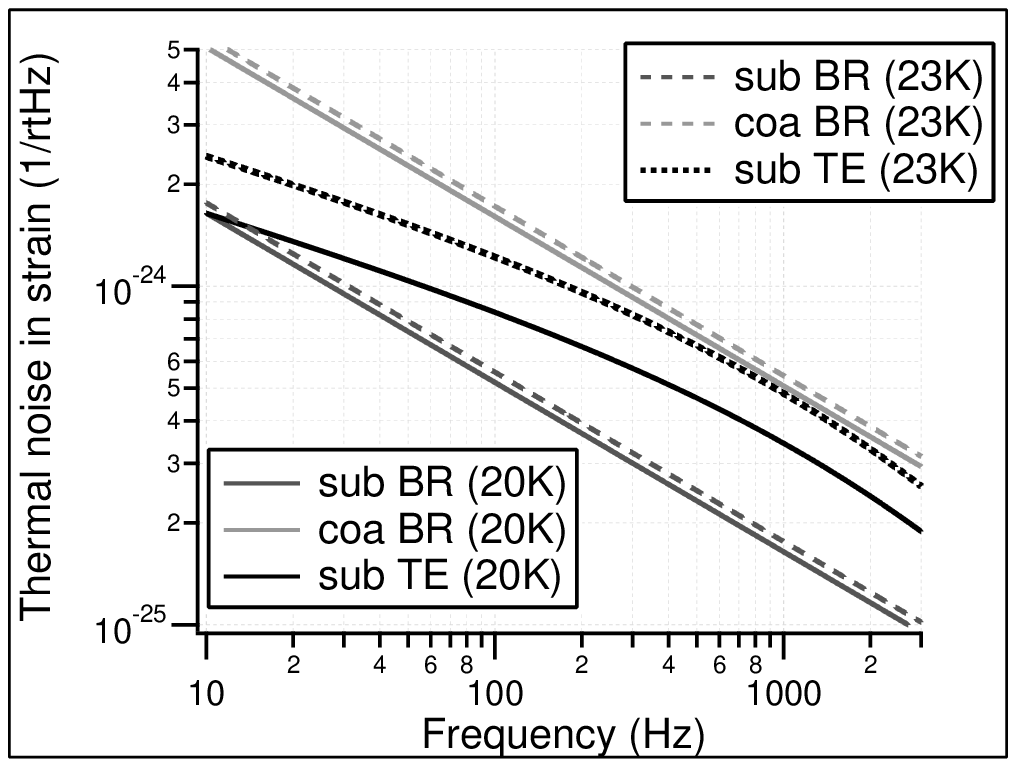}
		\includegraphics[width=7cm]{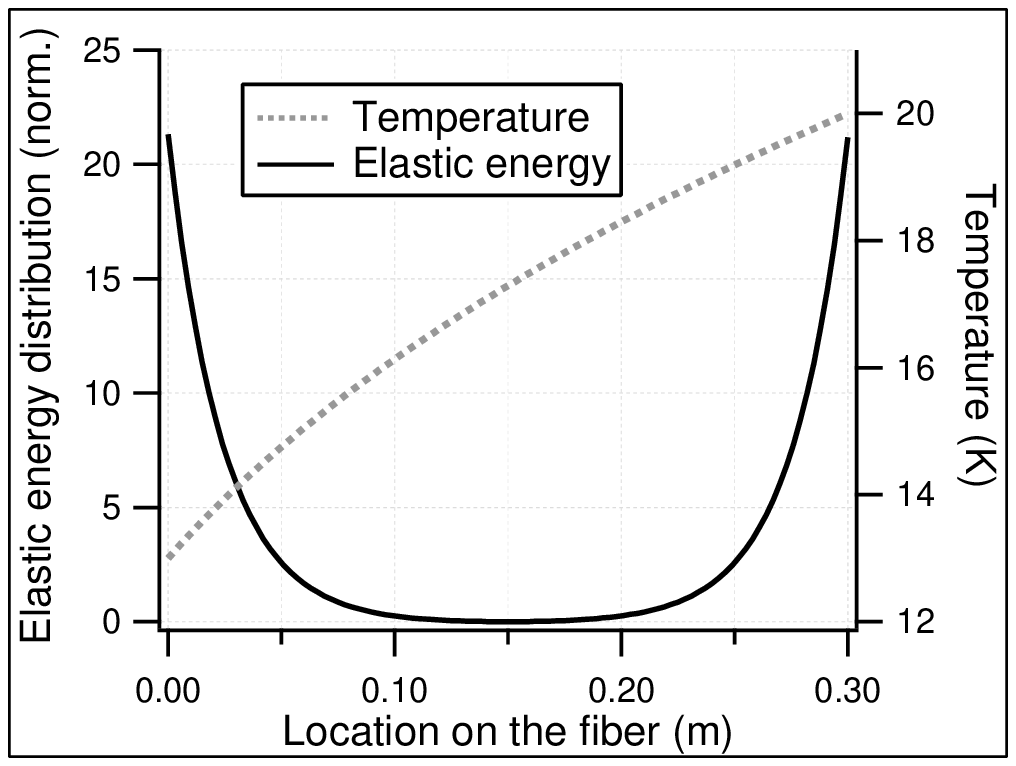}
	\caption{{\it Left}: Temperature dependence of mirror thermal noise. {\it Right}: Dissipation and temperature profiles of a suspension fiber.}
	\label{fig:TN}
	\end{center}
\end{figure}

Substrate thermoelastic noise at an arbitrary temperature is derived by Cerdonio and the low-temperature approximation is introduced by Yamamoto~{\cite{Cerdonio}\cite{Remark}}:
\begin{eqnarray}
&&S_\mathrm{TE}(\Omega)=\frac{4k_\mathrm{B}T^2(1+\nu_\mathrm{s})^2\alpha_\mathrm{s}^2}{\sqrt{\pi\kappa_\mathrm{s}C_\mathrm{s}\Omega}}\ \ \ (\mathrm{low\ temperature};\ \Omega\ll2\kappa_\mathrm{s}/C_\mathrm{s}w_0^2)\ ,\label{eq:TE}
\end{eqnarray}
where $\alpha_\mathrm{s}$, $\kappa_\mathrm{s}$, and $C_\mathrm{s}$ are thermal expansion rate, thermal conductivity, and specific heat per volume, respectively. Note that the thermoelastic noise level at low temperature does not depend on the beam radius. The thermal parameters $\alpha_\mathrm{s}$, $\kappa_\mathrm{s}$, and $C_\mathrm{s}$ depend highly on the mirror temperature~\cite{parameters}, for which thermoelastic noise increases remarkably with the mirror temperature as is shown in the left panel of Fig.~\ref{fig:TN}.

Reduction of mirror thermal noise by decreasing the temperature is observed at CLIO, a 100-m prototype interferometer for KAGRA. The sapphire ITMs in the arm cavities are cooled down to $\sim20$~K and the sensitivity is improved at around $100\sim200$~Hz~\cite{CLIO}.

\subsection{Suspension thermal noise}

Heat absorbed in the test mass is transferred to the intermediate mass through the sapphire suspension fibers. In order to extract a sufficient amount of heat, the fibers have to be as thick as 1.6~mm in diameter. The suspension thermal noise level at low frequencies is roughly given by the following equation~\cite{Saulson}:
\begin{eqnarray}
S_\mathrm{sus}(\Omega)\approx\frac{4k_\mathrm{B}T}{m\Omega^5}\sqrt{\frac{4\pi Y_wg}{m}}\left(\frac{d_w}{\ell_\mathrm{sus}}\right)^2\phi_w\ ,
\end{eqnarray}
with $g$, $Y_w$, and $\phi_w$ being gravity acceleration, Young's modulus of fiber, and mechanical loss of fiber, respectively, and one can see that the effect of low $T$ is compensated by large $d_w$.

For the actual noise calculation, we use the model developed for Virgo~\cite{PPP}. The model treats the 3-mass system and the dissipations of the intermediate mass and the recoil mass are included. In KAGRA, the intermediate mass is planned to be suspended by tungsten fibers and the recoil mass is planned to be suspended by BeCu fibers. Though the temperature of these masses is lower than that of the test mass, the mechanical losses of the tungsten and BeCu fibers are higher than the loss of sapphire, so that the dissipations of these fibers dominate suspension thermal noise at frequencies below 30~Hz.

The dissipation of the pendulum is mainly at the top and bottom ends of the fiber (see the right panel of Fig.~\ref{fig:TN}). The effective temperature to calculate suspension thermal noise in the horizontal mode is then the average of the test mass temperature and the intermediate mass temperature. Besides, contribution of thermal noise in the vertical mode is non-trivial and the effective temperature for this mode is the average over the fiber according to the temperature profile.

The peak of suspension thermal noise at 130~Hz is the vertical mode resonance and the peak at 230~Hz is the first violin mode. These peaks standing in the middle of the observation band can be troublesome as the peaks can be broadened by couplings to other noise sources such as seismic noise or laser intensity noise. However, the vertical resonance is inversely proportional and the first violin mode frequency is proportional to a square root of the fiber cross section, and vice versa to the fiber length, so that there is no solution to remove both peaks from the observation band by changing the fiber dimensions. A study to use a non-uniform fiber is under way.

\section{Quantum non-demolition techniques}\label{sec:QN}

Beating the standard quantum limit (SQL) in KAGRA is important for two reasons. One is that the laser power is lower than other advanced detectors. Absence of the thermal lensing problem will let us increase the laser power more smoothly, but we will have to stop increasing the power at a certain point when thermal noise starts limiting the sensitivity. With the limited power, a quantum non-demolition (QND) technique is a way to further improve the sensitivity. The other reason is that the expected gain of the QND technique is high with the low thermal noise level. The control scheme may become slightly complicated to implement QND techniques but it is worth putting in some extra efforts.

We use two QND techniques. The first one is back-action evasion (BAE). Let us assume that we operate the interferometer with zero detuning. With the conventional readout in the phase quadrature, shot noise that decreases with the laser power and quantum radiation pressure noise that increases with the laser power are not correlated so that the sensitivity does not exceed a limit by simply changing the power:
\begin{eqnarray}
&&S_\mathrm{sh}\ =\ \left(\frac{1}{4{\cal F}}\right)^{\!\!2}\cdot\frac{\hbar\lambda\pi c}{I_\mathrm{RSE}}\cdot\frac{\gamma^2+\Omega^2}{\gamma^2}\ ,\ \ \ 
S_\mathrm{rp}\ =\ \left(\frac{4{\cal F}}{m\Omega^2}\right)^{\!\!2}\cdot\frac{\hbar I_\mathrm{RSE}}{\lambda\pi c}\cdot\frac{\gamma^2}{\gamma^2+\Omega^2}\ ,\nonumber\\
&&S_\mathrm{total}^\mathrm{conv}\ =\ S_\mathrm{sh}+S_\mathrm{rp}\ \geq\ S_\mathrm{sql}\ =\ \frac{2\hbar}{m\Omega^2}\ .\label{eq:SQL}
\end{eqnarray}
Here ${\cal F}$, $\hbar$, $\lambda$, $c$, and $\gamma/(2\pi)$ are finesse of the arm cavity, reduced Planck constant, wavelength, speed of light, and cavity pole of the arm cavity, respectively, and $I_\mathrm{RSE}$ is frequency-dependent effective laser power of the RSE system given by~\cite{BC}
\begin{eqnarray}
I_\mathrm{RSE}&=&\frac{I_\mathrm{BS}(1-r_\mathrm{s}^2)}{1+r_\mathrm{s}^2+2r_\mathrm{s}\cos{[2\arctan{(\Omega/\gamma)}]}}\ ,
\end{eqnarray}
with $I_\mathrm{BS}$ and $r_\mathrm{s}$ laser power at the beamsplitter and amplitude reflectivity of the signal recycling mirror, respectively. Note that radiation pressure noise in Eq.~(\ref{eq:SQL}) is for a single mirror, so the mass $m$ should be replaced by the reduced mass $m/4$ for the 4-mirror system. The BAE introduces a correlation between shot noise and radiation pressure noise, which is realized by changing the readout quadrature. Since the source of radiation pressure noise is quantum fluctuations in the amplitude quadrature, one can measure it together with the source of shot noise in the phase quadrature and compensate radiation pressure noise. The total quantum noise spectrum reads
\begin{eqnarray}
&&S_\mathrm{total}^\mathrm{BAE}\ =\ S_\mathrm{sh}+\left(\sqrt{S_\mathrm{rp}}+\sqrt{S_\mathrm{sh}}\cot{\zeta}\right)^2\ ,\label{eq:SQL2}
\end{eqnarray}
which can be lower than $S_\mathrm{sql}$ at around a certain frequency depending on the readout quadrature $\zeta$.

\begin{figure}[t]
	\begin{center}
		\includegraphics[width=6.5cm]{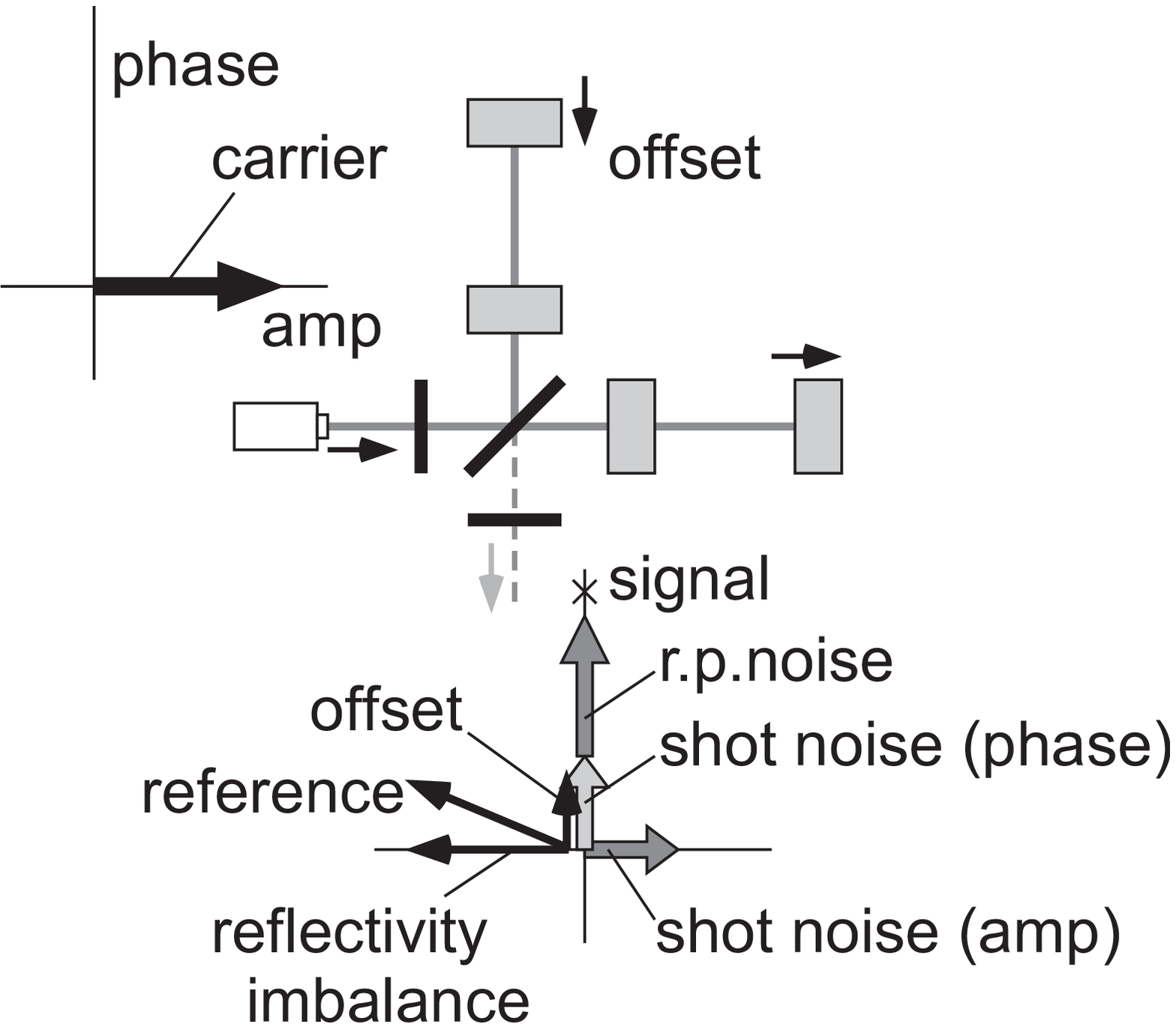}
\hspace{0.5cm}
		\includegraphics[width=7cm]{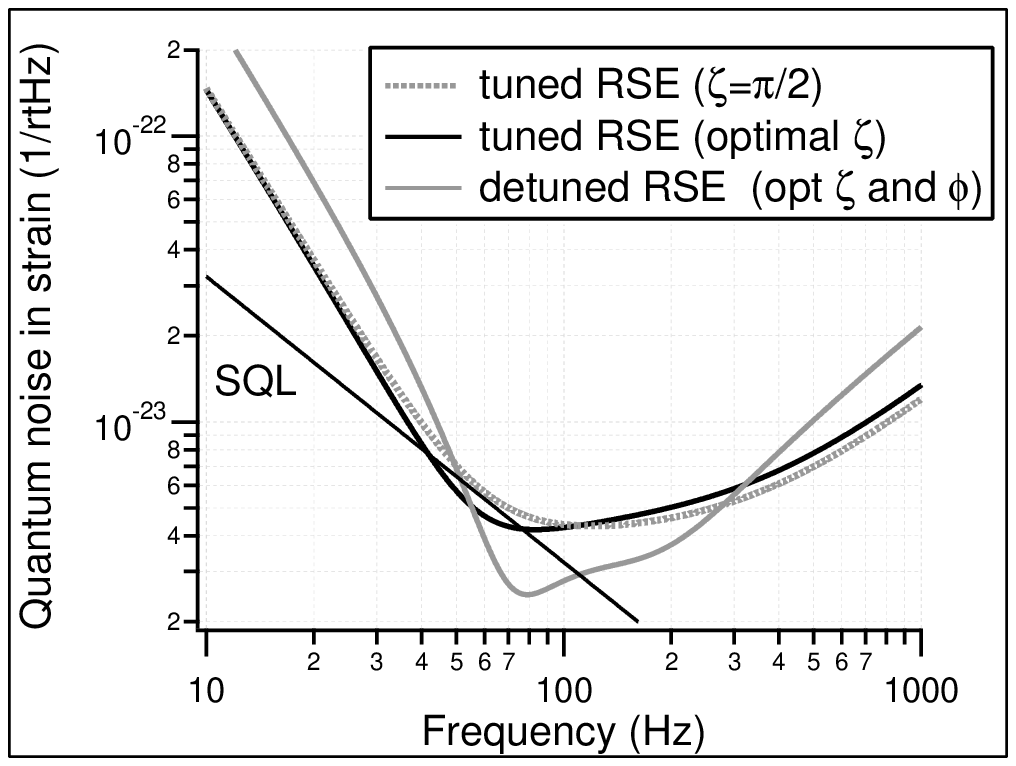}
	\caption{{\it Left}: Schematic view of the BAE readout. The output of the interferometer contains reference light, signal, and quantum fluctuation fields. Adjusting the offset to choose a proper readout quadrature, one can compensate radiation pressure noise (r.p. noise) with shot noise in the amplitude quadrature. {\it Right}: Quantum noise spectra of the variable RSE system in different operations.}
	\label{fig:BAE}
	\end{center}
\end{figure}

The left panel of Fig.~\ref{fig:BAE} explains how one can change the readout quadrature. The reference field that couples to gravitational-wave signals for the conventional measurement in the phase quadrature is the carrier light leaking from the arm cavity due to the intentional offset on the mirror position. This scheme is called the DC-readout scheme. There is also carrier light in the amplitude quadrature, which leaks through due to the reflectivity imbalance of the two arm cavities. Applying sufficient amount of the offset, one can measure the signal almost in the phase quadrature, but tuning the amount of the offset, one can choose the readout quadrature~\cite{DCreadout}. Since the reference light has to be weaker than that for the phase quadrature measurement, junk light must be well suppressed to realize the BAE readout.

The second QND technique is to use an optical spring, which can be realized by detuning the signal-recycling cavity~\cite{BC}. With the detuning, the gravitational-wave signal in the phase quadrature reflects back to the interferometer in an intermediate quadrature. The amplitude quadrature component couples with the laser light to produce radiation pressure on the test masses. The motion caused by the radiation pressure generates a signal in the phase quadrature. In the end there is a loop of the signal, which makes the optical spring, enhancing the susceptibility of the interferometer to the external force at a certain frequency depending on the laser power, mass, and the detune phase $\phi$. The equation to calculate the quantum noise level of the detuned interferometer is shown in Ref.~\cite{BC}; we shall omit to write it down here.

KAGRA will employ a so-called variable RSE system. The tuned configuration and the detuned configuration can be switched simply by changing the offset in the control loop of the signal recycling cavity. The BAE readout is applied with the tuned configuration and the optimal $\zeta$ to maximize the inspiral range for the neutron-star binaries is chosen. In the case of the detuned configuration, the optimal $\zeta$ and $\phi$ are chosen to maximize the inspiral range with the optical spring. In fact, a highly detuned configuration can increase control noise~\cite{Aso}. In the baseline design, the detune phase $\phi$ is set to 3.5~deg while the optimal one is 4.2~deg. The inspiral ranges of the various setups are shown in Table~\ref{tab:IR}. With the variable RSE system, we will start the operation in the detuned configuration to detect gravitational waves from neutron-star binaries, the primary target source, and then we can choose to extend the observation bandwidth with the tuned configuration.

\begin{table}[h]
\begin{center}
\begin{tabular}{|c|c|c|c|}
\hline
&$\phi$&$\zeta$&range\\ \hline
tuned RSE (conv.)&--&90~deg&196~Mpc\\ \hline
tuned RSE (BAE)&--&116~deg&206~Mpc\\ \hline
detuned RSE&3.5~deg&132~deg&238~Mpc\\ \hline
detuned RSE&4.2~deg&137~deg&239~Mpc\\ \hline
\end{tabular}
\caption{Inspiral ranges for neutron-star binaries with different setups.}
\label{tab:IR}
\end{center}
\end{table}

\begin{figure}[t]
	\begin{center}
		\includegraphics[width=10cm]{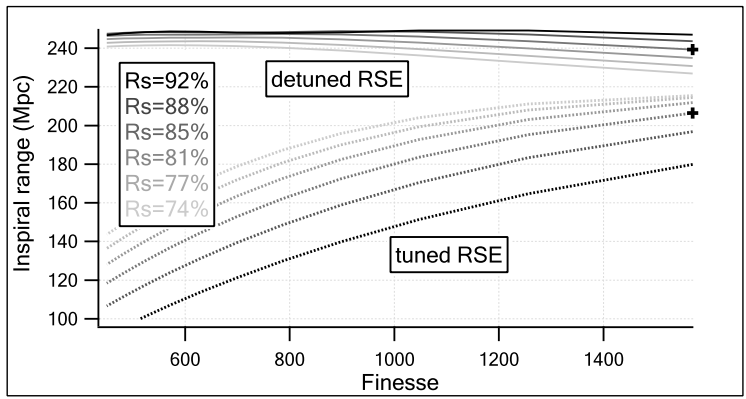}
	\caption{Inspiral range with different finesse and signal recycling mirror reflectivity.}
	\label{fig:IR}
	\end{center}
\end{figure}

The finesse of the arm cavities and the reflectivity of the signal recycling mirror are carefully chosen in terms of the bandwidth and the inspiral range. Figure~\ref{fig:IR} shows the inspiral range with different finesse and different reflectivity of the signal recycling mirror. The laser power is determined with a condition that the mirror temperature is 20~K. At each point, the readout quadrature and the detune phase (for the detuned configurations) are chosen to maximize the inspiral range. A tuned RSE and a detuned RSE with the same finesse and the same reflectivity are compatible. With the finesse being decreased, the range of detuned RSE increases up to a certain level that is determined by the mirror thermal noise level, but the range of tuned RSE decreases much faster so that we do not see much benefit in this direction. Besides, the optimal detune phase with the low finesse configuration is quite high and the bandwidth of the interferometer is narrow. This leads to degrade the estimate accuracy of the arrival time of gravitational waves. On the other hand, with the finesse and the signal-recycling mirror reflectivity increased too much, the optical loss in the signal recycling cavity degrades the sensitivity even though one can keep the mirror temperature to 20~K with more power in the arm cavities. In the end, we choose the parameter set indicated by the pair of $+$ marks: finesse of 1550 and signal recycling mirror reflectivity of 85~\%.

\section{Summary}

We introduced in this paper the configuration of the Japanese second-generation gravitational-wave detector KAGRA (previously called LCGT). With the underground site, the cryogenic operation, and the quantum non-demolition techniques, KAGRA will realize its extremely high sensitivity and observe gravitational waves in the next $6\sim7$ years. Together with other advanced detectors in the world, we will open the window toward the gravitational-wave astronomy.

\appendix
\section{List of parameters}\label{appA}

\begin{table}[h]
\begin{center}
\begin{tabular}{c|c}
&\\ \hline
finesse&1550\\
test mass diameter&$22\sim25$~cm\\
test mass thickness&15~cm\\
mass&$22.8\sim30$~kg\\
laser power at the beamsplitter&$515\sim825$~W\\
mirror temperature&20~K\\
absorption in ITM substrate&$20\sim50$~ppm/cm\\
absorption in coatings&$0.5\sim1.0$~ppm\\
optical loss of a test mass&45~ppm\\
transmittance of ETM&10~ppm\\
ITM/ETM radii of curvature&1680/1870~m\\
beam radii on ITM/ETM&3.5/4.0~cm\\
power/signal recycling mirror reflectivity&90/85~\%\\
sapphire fiber length&30~cm\\
sapphire fiber diameter&1.6~mm\\
tunnel tilt&1/300\\
vertical-horizontal coupling&1/200\\
mechanical loss of substrate&1e-8\\
mechanical loss of silica/tantala coatings&3e-4/5e-4\\
mechanical loss of fiber&2e-7\\
vacuum level&2e-7~Pa\\
\end{tabular}
\caption{Parameter list of the baseline-design KAGRA.}
\label{tab:list}
\end{center}
\end{table}


\begin{thebibliography}{10}
\bibitem{parameters} Temperature dependence of each thermal parameter is given as follows, fitted to measurement results:
$\kappa_\mathrm{s}=(0.272\times T_\mathrm{m}^{-2.2}+6.22\times10^{-9}\times T_\mathrm{m}^{3.04})^{-1.25}$~[W/m/K],
$C_\mathrm{s}=0.227\times T_\mathrm{m}^{3.14}$~[J/$\mathrm{m}^3$/K],
$\alpha=7.21\times10^{-13}\times T_\mathrm{m}^{2.99}$~[1/K], where $T_\mathrm{m}$ is in Kelvin.
\end{thebibliography}
\end{document}